\magnification=\magstep1

\hsize=6.5truein
\hoffset=-0.00truein
\baselineskip=16truept plus 0 truept minus 0 truept
\overfullrule=0pt
\tolerance=10000

\newcount\notecnt
\notecnt=0
\def\adnote{\advance\notecnt by1}
\def\fnote{\adnote \footnote{$^{[\number\notecnt]}$}}

\pageno=0 \bigskip 

\centerline{\bf Triton's Surface Age and Impactor Population Revisited in Light of Kuiper
Belt Fluxes:}

\centerline {\bf Evidence for Small Kuiper Belt Objects and Recent Geological Activity}

\bigskip
\bigskip
\bigskip

\bigskip

\centerline{S.~Alan Stern}

\centerline {and}

\centerline{William B.~McKinnon$^1$}
\bigskip
\bigskip
\bigskip
\centerline{Department of Space Studies}
\centerline{Southwest Research Institute}
\centerline{1050 Walnut Street, Suite 426}
\centerline{Boulder, CO 80302}
\centerline{astern@swri.edu, mckinnon@levee.wustl.edu}

\centerline { }
\centerline { }
\centerline { }
\centerline { }

\centerline { }
\centerline { }
\centerline { }
\centerline { }
\centerline { }

\centerline { }
\centerline { }
\centerline { }
\centerline { }
\centerline { }

\noindent 13 Pages

\noindent 02 Figures

\noindent 00 Tables


\bigskip
\noindent Submitted to {\it The Astronomical Journal:} 15 July 1999

\noindent Revised: 05 Oct 1999

\bigskip

\noindent $^1$On sabbatical from Department Earth and Planetary Sciences and McDonnell
Center for Space Sciences, Washington University, Saint Louis, MO 63130;
mckinnon@levee.wustl.edu

\vfill
\eject

\centerline {\bf ABSTRACT}

\medskip \noindent {\bf Neptune's largest satellite, Triton, is one of the most fascinating and
enigmatic bodies in the solar system.  Among its numerous interesting traits, Triton appears to
have far fewer craters than would be expected if its surface was primordial.  Here we combine
the best available crater count data for Triton with improved estimates of impact rates by
including the Kuiper Belt as a source of impactors.  We find that the population of impactors
creating the smallest observed craters on Triton must be sub-km in scale, and that this
small-impactor population can be best fit by a differential power-law size index near --3.  Such
results provide interesting, indirect probes of the unseen small body population of the Kuiper
Belt.  Based on the modern, Kuiper Belt and Oort Cloud impactor flux estimates, we also
recalculate estimated ages for several regions of Triton's surface imaged by Voyager 2, and find
that Triton was probably active on a time scale no greater than 0.1--0.3 Gyr ago (indicating
Triton was still active after some 90\% to 98\% of the age of the solar system), and perhaps
even more recently.  The time-averaged volumetric resurfacing rate on Triton implied by these
results, 0.01 km$^3$ yr$^{-1}$ or more, is likely second only to Io and Europa in the outer
solar system, and is within an order of magnitude of estimates for Venus and for the Earth's
intraplate zones.  This finding indicates that Triton likely remains a highly geologically
active world at present, some 4.5 Gyr after its formation.  We briefly speculate on how such a
situation might obtain.}

\bigskip


\vfill
\eject

\centerline {\bf 1.~INTRODUCTION}

\bigskip \noindent The 1989 Voyager 2 encounter with the largest satellite of Neptune,
Triton, revolutionized our knowledge of this world, revealing it to be a scientifically
inspiring satellite, 2700 km in diameter, with an N$_2$/CH$_4$ cryo-atmosphere, and a
morphologically complex surface (e.g., Smith et al.~1989).  Voyager also discovered detached
hazes, atmospheric emissions excited by the precipitation of charged particles from
Neptune's magnetosphere, and small vents generating plumes that rise almost 10 kilometers
through Triton's atmosphere.

\medskip \noindent Among the most intriguing questions concerning this distant world is the
issue of its surface age and therefore, by extension, the degree of recent or ongoing internal
activity within this body.  Voyager era investigators obtained a crude global surface age
estimate of 1 Gyr (Smith et al.~1989; Strom et al.~1990; cf.~Croft et al.~1995), but their
calculations did not take into account the cratering flux from the (then undiscovered) Kuiper
Belt.

\medskip \noindent In what follows we will combine existing crater counts with modern impact
flux estimates, which include the Kuiper Belt, in order to derive new estimates of the surface
age on Triton.  

\medskip \noindent This paper extends considerably some preliminary results reported in an LPSC
abstract (Stern \& McKinnon 1999).  A key assumption we make is that the primary process that
removes craters from Triton's surface is conventional geological activity (e.g., volcanism), as
opposed to a more exotic possibilities such as viscous relaxation, escape erosion, or charged
particle degradation.  This assumption is strongly supported by Triton's apparently conventional
crater size-frequency distribution, and the uniformly fresh appearance of craters on Triton.

\smallskip
\bigskip
\centerline {\bf 2.~ATTRIBUTES OF TRITON'S CRATER AND IMPACTOR POPULATIONS}
\bigskip

\noindent We begin by estimating the typical size scale for craters produced by impacting
bodies.  Following standard Schmidt-Holsapple crater scaling (e.g., Chapman \& McKinnon
1986; Holsapple 1993), the crater diameter $D$ for a specified set of impact parameters
and surface properties on a body with gravity $g$ can be estimated from:

$$
   D_{tr} = 1.56 d (A \delta/\rho)^{1/3}
            (1.61 g d/v^2)^{-\alpha/3} ({\rm cos}\bar{\theta})^{2\alpha/3}, \eqno (1)
$$

\smallskip \noindent where $D_{tr}$ is the so-called transient diameter, which we assume to be a
paraboloid of revolution with a depth/diameter ratio of 1/2$\sqrt{2}$ (McKinnon \& Schenk 1995).
Here $d$ is the equivalent spherical impactor diameter, $\delta$ and $\rho$ are the impactor and
surface densities, respectively, $v$ is the impact velocity, $A$ and $\alpha$ are scaling
constants which depend on the thermomechanical properties of the surface, and
cos$\bar{\theta}$=0.71 is an adjustment factor to account for the average impact angle
($\theta$=45 $\deg$).  We adopt a maximum impactor velocity $v_{max}$=11.6 km s$^{-1}$, as set by
the root-sum-square of Triton's escape speed and the sum of Triton's orbital speed and the maximum
impactor velocity at Triton's orbit.  We adopt a minimum impactor velocity $v_{min}$=2.3 km
s$^{-1}$, as set by the root-sum-square of Triton's escape speed and the difference between
Triton's orbital speed and the escape speed from Triton's orbit.

\medskip \noindent Now, $D_{tr}$ is proportional to the final crater diameter $D$ for $D$$<$$D_c$,
where $D_c$ is the simple-to-complex crater transition diameter, which is between 6 and 11 km on
Triton (Strom, Croft, \& Boyce 1990; Schenk 1992).  We take $D_c$=8 km.  When $D$$>$$D_c$, i.e., in
the case of complex (flattened) craters, the scaling relationship is also relatively
straightforward.  Based on both morphological measurements of craters on Triton (Croft et al.~1995),
and geometrical models of craters on Ganymede (Schenk 1991; McKinnon \& Schenk 1995), the closest
analogues to Triton's craters for which extensive data exist, we therefore write:

$$
D(D<D_c) = 1.3D_{tr}  \eqno (2a)
$$

$$
D(D>D_c) = (1.3D_{tr})^{1.11}D_c^{-0.11}.  \eqno (2b)
$$

\smallskip \noindent The scaling presented in Equations (1) and (2) are probably accurate to
30\% in $D$.  

\smallskip \noindent To fully explore the range of impactor diameter $d$ that generates the
observed craters on Triton, we show Figure 1. This figure evaluates Equations (1) and (2) as a function
of impactor diameter over both the range of probable impact velocities, and a suite of
($\delta$,$\rho$) cases spanning the reasonably expected parameter space.  The baseline case
against which other density pairs may be compared is simply the one of equal densities for
impactor and surface (Fig.~1, lower right).

\medskip \noindent Inspecting Figure 1, one concludes that the 2.8 to 27 km diameter craters
identified in Voyager images of Triton (Smith et al.~1989; Strom, Croft, \& Boyce 1990) imply
impactors with diameters between 0.1 km and 0.7 km (to create the 2.8 km minimum crater diameters
counted), and 2--11 km in diameter (to create the largest craters detected, like Mazomba with
$D$=27 km).  Such sizes naturally imply comet-sized bodies as the dominant observed Triton
impactor population, and as such provide a valuable constraint on the small body population in
Neptune's region of the solar system.  This is our first result.

\medskip \noindent We now consider the crater size and number statistics derived from imaging by the
Voyager 2 spacecraft during its 1989 flyby of Triton, in order to constrain the size-frequency power
law index of the impactor population.  The best available assessment of Triton's crater statistics
(Smith et al.~1989; Strom et al.~1990) discussed four regions on Triton (``Areas 1--4'') on which
careful crater counts were attempted.  These areas add up to 16\% of Triton's total surface area,
out of a total of $\approx$40\% of the satellite that was imaged by Voyager at resolutions useful
for geological analysis.  The interested reader can find images of Areas 1--4 and a sketch map
showing their location on Triton in Smith et al.~(1989).\fnote {We do not consider the results of
the 6 highest-resolution Voyager frames, reported in summary form by Croft et al.~(1995), because
the area covered is small and the images used are smeared to varying degrees owing to spacecraft
motion.}  Because the observed crater population on Triton in general, and Area 1 in particular, is
far from saturation equilibrium, and is not manifestly geologically degraded, the Triton crater
counts probably represent a production population.

\medskip \noindent We concentrate initially on Area 1, which exhibits the highest density of
craters and therefore has the best statistical confidence.  Area 1 is located near the apex of
Triton's orbital motion, and contains 9.79$\times$10$^5$ km$^2$ (some 4.2\% of Triton's surface
area); it displays a total of 181 craters with d$>$2.8 km.  We concur with Strom et al.~(1990)
that of the 4 distinct terrains counted on Triton, Area 1 has the crater count statistics to best
support a size-frequency analysis.

\medskip \noindent For the small-body impactor population we take a differential power law, as is
typically used to represent the Kuiper Belt population (Weissman \& Levison 1997), of the form
$n$($d$) $\propto$ $d^{b}$.  We again select impact velocities to evaluate Equation (1) from a uniform
velocity distribution between the probable $v_{min}$ to $v_{max}$ range for Triton, and use Equations
(1) and (2) to scale from impactor diameters to crater diameters.

\medskip \noindent Figure 2 shows the results of model simulations designed to fit $b$, the
power-law exponent on the size-frequency distribution of impactors in Area 1.  Figure 2 shows that
the Voyager data are fit well by relatively shallow power-law slopes of the impactor population,
with the nominal value of $b$ being near --2.5.  This result is robust to the choice of plausible
($\delta$,$\rho$) combinations in Equation because this ratio appears as a multiplier in the scaling
equation and the change in slope at $D$=$D_c$ is not severe.  We note, however, that the resolution
in some of the images used in the Area 1 count are as poor as 2.2 km line-pair$^{-1}$; therefore the
bottom-most bin or two likely suffered undercounting.  Neglecting these bottom-most bin or two
allows steeper fits, up to $b$$\approx$--3, within the Poisson statistics of the crater counts.
In what follows we adopt $b$=--3 as our preferred solution.

\medskip \noindent The slope parameter just derived is in accord with both Weissman \& Levison's
(1997) Kuiper Belt model, and is also consistent with Shoemaker \& Wolfe's (1982) preferred --3.0
power-law index for ray-crater impactors on Ganymede (presumably comets).  This second result provides
a new (if indirect) source of information on the population of small bodies that cannot as yet be
optically detected in the Kuiper Belt, and indicates they are plentiful, as collisional evolution
models have predicted (e.g., Stern 1995, Davis \& Farinella 1997).

\smallskip \bigskip

\centerline {\bf 3.~IMPACTOR FLUXES}

\bigskip \noindent In preparation for estimating surface unit ages, we now estimate the
current cratering {\it rate} on Triton, $\dot{N}$.  The heliocentric flux contributing to
$\dot{N}$ consists of terms due to objects on Neptune-crossing orbits from both the
Edgeworth Kuiper Belt (EKB) (Levison \& Duncan 1997; hereafter LD97) and the Scattered
Kuiper Belt (SKB) (Duncan \& Levison 1997), and due to objects in the Oort Cloud (Weissman
\& Stern 1994). For a recent Kuiper Belt review, see Farinella et al.~(2000).

\medskip \noindent We neglect the possibility of a significant Neptuneocentric population of
impactors (Croft et al.~1995) on two grounds.  The first is the great observed emptiness of the
Neptunian system with regard to debris and small satellites outside 5 Neptune radii (Smith et
al.~1989).  The second is the fact, easily shown, that any small-body impactor population large
enough to populate Triton's surface with the Voyager-observed craters would, if their orbits are
Triton crossing, be swept up on time scales of 1 to 10$^3$ years in most cases.  Therefore, unless
a discrete Neptuneocentric flux event very recently populated Triton with its observed craters,
this short sweep up time for Neptuneocentric debris implies a large unseen population of such
impactors on Triton-crossing orbits.  Indeed, to sustain this population against Triton sweep up
over 100 Myr would imply both an accreted veneer of mass up to 10$^{24}$ gm (roughly the mass of a
typical Uranian satellite), {\it and} a surface that is constantly renewed on timescales of 10$^3$
years or less.   

\medskip \noindent We will refer to the combined EKB+SKB flux term as the KB contribution.  To
obtain the total Kuiper Belt cratering rate on Triton, we adopt the state-of-the-art comet impact
rate estimate by LD97, as revised by Levison et al.~(1999; henceforth LDZD99), i.e.,
$\dot{N}_{Neptune}=$3.5$\times$10$^{-4}$ comets yr$^{-1}$ with $d>2^{+2}_{-1}$ km on Neptune.  We
then scale that result to Triton, accounting for its smaller diameter and the gravitational focusing
at its distance from Neptune.  For an average encounter velocity at Neptune's sphere of influence of
$\approx$0.3 Neptune's orbital speed (LD97), these factors together conspire to reduce Triton's
collision cross-section, and therefore its globally-averaged collision rate, by a factor near
2.7$\times$10$^{-4}$, relative to Neptune.  Combined with the fact that the time-averaged Kuiper
Belt source rate into the planetary region has probably only declined by $\sim$5\% over the past 0.5
Gyr (Holman \& Wisdom 1993; Levison \& Duncan 1993), we predict a present-epoch, {\it globally
averaged} KB-impactor source rate of $\dot{N}_{KB}$=1.0$\times$10$^{-7}$ comets yr$^{-1}$ with
$d$$>$2 km, or, in a more useful, surface area-normalized form for our purposes,
$\hat{\dot{N}}$=4$\times$10$^{-15}$ craters km$^{-2}$ yr$^{-1}$ due to comets with d$>$2 km.  LD97's
estimated uncertainty in deriving the KB term for $\dot{N}_{Neptune}$ is of order a factor of 2.8 to
4, depending on whether the diameter uncertainty above is convolved with a $b$=--2.5 or $b$=--3
differential power-law size index, respectively.

\medskip \noindent Neither LD97 nor LDZD99 included an estimate of the Oort Cloud (OC) impactor
rate on Neptune.  Weissman \& Stern (1994), however, made a calculation for the Oort Cloud
impact rates on Pluto, an outer solar system object of similar physical cross-section to Triton
in a region with relatively similar OC flux.  They estimated that the total number of Oort Cloud
impacts by comets with $d$$>$2.4 km on Pluto is $\sim$50 over the past 4 Gyr.  Scaling this
result to Triton's larger physical cross-section and enhanced gravitational focusing environment
around Neptune, and then adopting a (limiting case) --3 differential power-law size index, we
expect $\sim$150 impacts with $d>$2 km on Triton over the past 4 Gyr.  This corresponds to an
average impact rate of $\dot{N}_{OC}$=3.7$\times$10$^{-8}$ yr$^{-1}$ for $d>$2 km, or
$\hat{\dot{N}}$=1.6$\times$10$^{-15}$ craters km$^{-2}$ yr$^{-1}$ due to comets with $d>$2 km,
some $\approx$40\% of the d$>$2 km KB impactor rate.  This value, 40\%, is likely to be an upper
limit, however, because it assumes the perihelion distribution of inner Oort Cloud comets
extends smoothly across Neptune's dynamical barrier, which is unlikely (Weissman \& Stern 1994).
This leads us to conclude that the EKB+SKB flux is clearly the dominant contributor to recent
cratering on Triton, our third result.

\medskip \noindent Because the OC cratering rate appears to be only $\sim$40\% or less of the
EKB+SKB cratering rate, we neglect it in what follows.

\medskip \noindent Continuing, for a satellite in synchronous rotation like Triton, it is well
known that the area around the apex of motion is where impact fluxes should be highest
(Shoemaker \& Wolfe 1982).  And indeed as noted above, Area 1, which is near the apex of
motion, was the most heavily cratered terrain seen on Triton.  Therefore, we must account for
this position-dependent flux in interpreting unit ages where craters have been counted on
Triton.

\medskip \noindent Shoemaker \& Wolfe (1982) showed that the enhancement factor, $\eta_1$, for any
given surface unit $i$, is close to a factor of 2 near the apex, and varies approximately as the
cosine of the angular distance from the apex of motion, reaching unity 90 deg from the apex (see
their Eqn.~17).  Area 1 stretches from about 20 to 60 deg from the apex of motion, which yields an
area-averaged factor of $\eta_1$=1.8 increase for Area 1 in its nominal cratering rate over that
predicted for the global-average Triton.

\smallskip 
\bigskip 
\centerline {\bf 4.~CRATER RETENTION AGES} 
\bigskip

\noindent The average crater retention age of Area i's surface can be estimated from the
general relation:

$$
   T_i = \left( {{N_{{\rm crat},i}} \over {\hat{\dot{N}}}} \right)
         \left( {{1} \over {\eta_i A_i}} \right). 
         \eqno (3)
$$

\smallskip \noindent Here $\eta_i$ is from above, $A_i$ is the area of the unit and $N_{{\rm
crat},i}$ is the number of craters formed on that unit by impactors of $d>$2 km; recall
$\hat{\dot{N}}$ is a global average for Triton.

\medskip \noindent We now evaluate Equation (3) for Area 1, where $\eta_1$=1.8.  If we presume
that comets are relatively dense and Triton's surface is no denser than pure water ice (Fig.~2,
lower left), then the age computed from Eqn.~(3) assuminng b=--3, is 240 Myr (600 Myr for
b=--2.5).  More plausibly, for the baseline case of equal densities for impactor and surface, we
find $T_1$=320 Myr (750 Myr for b=--2.5)\fnote {We base the Area 1 age determination for b=--3
on the 99 craters with D$>$4 km counted by Strom and coworkers.}.  The key implication, which is
robust for most plausible combinations of $\delta$ and $\rho$, is that Area 1 is geologically
very young, almost certainly $<$10\% of the age of the solar system, and perhaps a good deal
younger.

\medskip \noindent These estimates for $T_1$ likely represent an {\it upper limit} on the time
scale for the most recent significant geologic activity on Triton.  Why?  Even for the limited
fraction of Triton imaged at decent resolution by Voyager, stratigraphic relationships show that
Area 1 is older than adjacent units on Triton (Croft et al.~1995), particularly Smith et al.'s
(1989) Areas 2 and 4.  Area 2 is an assemblage of volcanic plains, and is about half as densely
cratered at Area 1 (Strom et al.~1990).  Because Area 2 stretches from 60$^\circ$ to 90$^\circ$
from the apex of motion, $\eta_2$=1.25, which yields a baseline (i.e., $\rho$=$\delta$) $T_2$ of
$\approx$ 230 Myr (550 Myr for b=--2.5); as above, these estimates assume all of these craters
are due to sources outside the Neptune system.  Areas 2 and 4 have similar crater densities and
lie at similar distances from the apex of motion, so $T_4$ is similar to $T_2$.  We note,
however, that Area 4 comprises part of the northern portion of Triton's southern frost cap, and
is thus subject to a variety of surfacial modification processes, so the formation age of this
unit may in fact be somewhat older.  The craters initially identified on Area 3, Triton's
cantaloupe terrain, were later shown unlikely to be due to impact (Strom et al.~1990).

\medskip \noindent Taken together, it is clear that three of the four crater-mapped units on
Triton yield crater retention ages that are not only substantially less than 1 Gyr, but may well
be of order 0.2--0.3 Gyr.  This is our fourth result.\fnote {And why are there no very large
impact craters or basins on Triton?  Because Triton's surface is too young to preserve the rare
impact scars from 100 km impactors.  A simple calculation of KB flux indicates that such objects
should impact Triton on timescales of $>$10$^{10}$ years.  (Regarding ancient impacts, the
thermal pulse associated with tidal breaking should have erased any primordial surface.)} The
primary reason for the younger ages we have just derived is the inclusion of the Kuiper Belt
population and its consequent effect on impact rates.

\medskip \noindent What factors could conspire to substantially increase our estimates of these
ages?  They could be increased if either $N_{{\rm crat},i}$ were larger, or $\dot{N}$ were
smaller.  However, because the crater counts are complete at large sizes, it is unlikely that
$N_{{\rm crat},i}$ can be substantially increased, particularly for Area 1, the oldest of the 4
units (with the most statistically robust crater counts).  Of course, an undercount could have
occurred if viscous relaxation or escape erosion has removed significant numbers of craters over
time.  But because the Strom et al.~(1990) crater counts rely only on fresh craters, and neglect
degraded ones (of which there are few if any known, a fact which itself argues for a recent
resurfacing), we believe that it is unlikely this is an important factor.  As for reducing
$\dot{N}$, there is the caveat noted above that LD97's impact flux (and that of LDZD99) carries
an estimated factor of $\sim$4 uncertainty, which could allow $T$ to exceed a Gyr; however, this
uncertainty is equally likely to {\it decrease} $\dot{N}$.  Another alternative would be if
Triton has until recently had a massive, impact shielding atmosphere (Lunine \& Nolan 1992);
however, there is no evidence for this in Triton crater morphologies or size-frequency
statistics.\fnote {Reported underabundances of small craters on the Galilean satellites (Chapman
et al.~1998) only occur at crater sizes ($D$$<$1 km) and for processes well below Voyager
resolution at Triton.}  We thus conclude that our estimated ages are unlikely to be
underestimated.

\medskip \noindent In contrast to the difficulty of raising $T$, it is easy to imagine lowering
Triton's crater retention age below our nominal estimates.  For example, as noted above, there
must have been some contribution to $\dot{N}$ from the Oort Cloud.  $T$ could also be lowered
if some fraction of the crater counts were due to other sources, such as:  (i) if impacting
populations other than the Kuiper Belt and Oort Cloud (e.g., Neptuneocentric) dominate, (ii) if
many of the craters counted are secondaries from larger craters on the unimaged parts of Triton,
or (iii) if endogenic (i.e., geological) processes, rather than impacts, created many of the
observed craters.  Concerning the first possibility, we have already argued above against a
dominant Neptuneocentric impactor population being very likely, but it is possible that, for
example, a fortuitous, recent Oort Cloud shower of significant magnitude could have produced a
cratering spike.  The latter two possibilities are also unlikely, as the identification of
impact crater morphologies on Triton's plains units (which includes Area 1) is generally clear,
and secondary crater populations characteristically follow steeper size-frequency distributions
(e.g., Melosh 1989).

\medskip \noindent A more serious matter concerns the overall KB cratering rate.  LDZD99's
revision of LD97 included both longer integration times (for better averages) and a comparison
of computed impact rates (direct counts) with those estimated by means of \"Opik's equations
from the modeled ensemble of ecliptic comets (for comet terminology, see Levison [1997]).  This
resulted in a factor of $\approx$3.5 reduction in impact rates relative to LD97.  LDZD99's new
impact rate estimates can be turned into cratering rates by calibrating the modeled comet
population against active, visible ecliptic comets and estimating the lifetime of the activity
(which yields the ratio of active to extinct comets), and estimating a minimum diameter (and
mass) for visible comets.  Such a procedure was in fact exploited by Zahnle et al.~(1998) in
their systematic study of cratering rates on the Galilean satellites.  In this work Zahnle et
al.~(1998) estimated that bombardment in the jovian system is dominated by Jupiter-family,
ecliptic comets (JFCs), both active and extinct, and at a rate lower than but within a factor
of two of that estimated by Shoemaker (1996).  Shoemaker's estimate, obtained using \"Opik's
equations, was dominated by extinct JFCs, and was based on an {\it observed population} of
asteroidal bodies in JFC-like orbits.  

\medskip \noindent \medskip \noindent The problem is this:  if the Zahnle, Dones, \& Levison
(1998) estimate is recalibrated to LDZD99, then their cratering rates on the Galilean satellites
fall by 3.5 and become $\approx$6 times less than Shoemaker's (1996) estimate for extinct JFCs
alone.  We are skeptical that Shoemaker's cratering rates are overestimated to such a degree,
especially as the logical chain from observed asteroid orbits and magnitudes to crater
production rates on the Galilean satellites is a short one.  There are cratering rate estimates
that are, conversely, much lower than even LDZD99 (i.e., Neukum et al.~1998), but these are
based on the assumption that the Gilgamesh basin on Ganymede is the same age as the Orientale
basin on the Moon, and otherwise ignore observations of present-day comets and asteroids.  We
discount these latter estimates.  Our view is that Shoemaker's estimates indicate that the
calibration in LDZD99 is probably low, and that the true cratering rate may be higher than we
obtained above by a sactor of up to $\approx$6.  If so, then all of the terrain ages derived above may
also be overestimates by a factor of several.  In particular, the age of Triton's leading
hemisphere plains (Area 1) may be of order 50 Myr, and the age of the young volcanic
plains on Triton (Area 2) may of order 40 Myr.

\smallskip \bigskip 

\centerline {\bf 5.~DISCUSSION}

\bigskip \noindent What are the implications of our results vis-\`a-vis Triton's activity?  To
begin, let us consider the time-averaged volumetric resurfacing rate on Triton, $\dot{V}_{TR}$.
A characteristic depth of several 100 m is required to overtop the rims of the largest craters
seen on Triton, and indeed, to bury most of the topographic structures observed (Croft et
al.~1995).  We assume a global resurfacing depth of 100 m over a
(conservative) timescale of 300 Myr, which gives a characteristic volumetric resurfacing rate on
Triton of $\dot{V}_{TR}$=0.01 km$^3$ yr$^{-1}$ (2.5$\times$10$^8$ yrs/$T$).  Based on
uncertainties in the required resurfacing depth and $T$, it is not implausible that the actual
value of $\dot{V}_{TR}$ has been or is a factor of several times higher.  Regardless, this
resurfacing rate is far higher than what can be supported by the small-scale plume vents seen by
Voyager, and indicates a far more active world in the geologically recent past than has been
formerly appreciated.

\medskip \noindent This conservative, 0.01 km$^3$ yr$^{-1}$ resurfacing rate also exceeds the
escape-loss erosion (Strobel \& Summers 1995) and aeolian transport (Yelle et al.~1995) rates on
Triton by two orders of magnitude.  We therefore conclude that geologic processes are indeed the
dominant surface modification process operating on a global scale on Triton (Croft et al.~1995).

\medskip \noindent Now consider Triton's volumetric resurfacing rate, $\dot{V}_{TR}$, in
comparison to other bodies.  The lunar resurfacing rate during the active, mare-filling
epoch was also $\sim$0.01 km$^3$ yr$^{-1}$ (Head et al.~1992).  The current-epoch volcanic
resurfacing rates on the Earth (Head et al.~1992), Venus (Bullock et al.~1993; Basilevsky et
al.~1997) and Io (Spencer \& Schneider 1996) are estimated to be $\approx$4 km$^3$
yr$^{-1}$, $\approx$0.1--0.4 km$^3$ yr$^{-1}$, and $\approx$40 km yr$^{-3}$, respectively;
the terrestrial rate excluding plate boundaries is $\approx$0.3--0.5 km$^3$ yr$^{-1}$ (Head
et al.~1992).  These various comparisons show that Triton {\it clearly} appears to be more
active than any other solid body in the outer solar system, except the tidally heated
satellites Io and Europa.\fnote {We must note that Titan's volumetric resurfacing rate is
unknown at present due to its opaque atmosphere.}

\medskip \noindent If Triton has been substantially internally active in the geologically
recent past, it is natural to imagine that Triton is still active today (or else the
geologic engine would have just run out, causing the Voyager observations to have occurred
at a ``special time'').  We therefore now consider the question of how Triton, which lacks
any significant present-day tidal forcing and has a radius of just 1350 km, could maintain
geologic activity 4.5 Gyr after its formation.  We discuss two ways which this could have
occurred.

\medskip \noindent First, Triton's own internal engine, powered by radiogenic energy
release alone, may after 4.5 Gyr still generate mantle temperatures exceeding 200 K
(Stevenson \& Gahndi 1990; McKinnon, Lunine, \& Banfield 1995).  Such conditions are
possibly significant enough to power widespread, low-temperature cryovolcanism that
accounts for the recent resurfacing.  This cryovolcanism could in principle also
be related to Triton's observed plume vents (Kirk et al.~1995) and global color changes
(Buratti et al.~1999). 

\medskip \noindent Alternatively, it is possible that Triton's recent geologic activity is instead
due to residual tidal heat resulting from a late-epoch capture into Neptunian orbit.  This scenario
would imply that Triton was a resident of the EKB or SKB until relatively recently (i.e., within the
last Gyr), and as discussed above, that it likely would then have possessed an atmosphere until even
more recently.  This scenario would in turn favor Triton's capture by collision with an original
satellite (Goldreich et al.~1989), rather than by means of gas drag in a proto-Neptunian nebula
(McKinnon \& Leith 1995).  Nevertheless, the {\it a priori} likelihood of such an event late in
solar system history, and from such a depleted reservoir as the present-day EKB or SKB, is low.

\medskip \noindent While a late capture is not impossible, it might seem simpler to
accept that Triton is big enough, and composed of mobile enough ices, to be geologically
active, as in the first scenario sketched above.  Given the accumulating evidence for warmth
and activity inside the Galilean satellites as revealed by the Galileo mission (e.g.,
McKinnon 1997), this should not be seen as so surprising.  Perhaps Triton is telling us that
somewhat smaller icy bodies, such as Pluto, can also remain geologically active at late times.

\smallskip \bigskip 

\centerline {\bf 6.~CONCLUSIONS}

\bigskip \noindent Combining Voyager-derived Triton crater counts and improved cratering
flux estimates for Triton, we have derived the following findings:

\medskip 1. The impactor population reaching Triton today is most probably dominated
by the Kuiper Belt. 

\medskip 2. Triton's extant surface craters require an impactor flux that contains a
substantial population of sub-km impactors; plausible 0.1 km to 10 km impactor populations
appear to exhibit power-law slope size indices in the range --2.5 to --3, with the steeper slope
being more likely.

\medskip 3. Findings 1 and 2 together imply a strong (if circumstantial) case for a
significant, unseen population of km-scale and sub-km scale bodies in the Kuiper Belt, as
predicted by both dynamical and collisional models (see Weissman \& Levison 1997; Farinella,
Davis, \& Stern 2000).

\medskip 4.  Unless the areas of Triton imaged by Voyager are not representative of the
object as a whole, Triton's global average surface age may be of order 100 Myr,
though older ages cannot be formally ruled out.  Regardless, this implies surface ages
for the imaged units that are at least a factor of 2, and perhaps over a factor of 10,
younger than the 1 Gyr derived at the time of the Voyager flyby (Smith et al.~1989).  Even
if unimaged terrains are more heavily cratered than the terrains seen by Voyager, the units
already mapped indicate very recent resurfacing  over large regions of Triton.

\medskip 5. As such, Triton appears to have been active throughout at least 90\%, and
perhaps over 98\%, of the age of the solar system.  These estimates are conservative, in
that some dateable units on Triton may well be significantly less than 100 Myr old.  It is
plausible that Triton's internal engine {\it still} supports sufficient ongoing activity
capable of generating large-scale (perhaps episodic) resurfacing.

\medskip 6. Triton's high rate of resurfacing may indicate its capture and subsequent,
tidally-driven thermal catastrophe occurred relatively recently; alternatively, the
high rate of resurfacing may imply that we understand less than had been thought about 
the interiors of icy objects like Triton and Pluto.

\medskip 7. Owing to the derived time-average volumetric resurfacing rate, $\dot{V}_{TR}$,
exceeds 0.01 km$^3$ yr$^{-1}$, geologic processes are clearly the dominant large-scale surface
modification process operating on Triton.

\medskip 8. Triton's inferred volumetric resurfacing rate exceeds all other satellites in
the solar system except Io, Europa, and possibly Titan (whose rate is unknown).  The
time-average resurfacing rate at late epoch is comparable to or exceeds the lunar 
resurfacing rate during the  Moon's active, mare-filling era.

\vfill \eject

\smallskip 
\bigskip 

\centerline { }
\centerline {\bf ACKNOWLEDGEMENTS}

\bigskip \noindent We thank Mark Bullock, Clark Chapman, Dan Durda, Hal Levison, Paul Schenk,
and Bob Strom for useful discussions, and Kevin Zahnle for comments on an earlier version of
this manuscript.  We further acknowledge helpful comments from an anonymous referee.  This
research was supported by the NASA Origins of Solar Systems (SAS, WBM) and the NASA Planetary
Geology and Geophysics (WBM) programs.

\vfill 
\eject

\centerline {\bf REFERENCES}
\bigskip

\noindent Basilevsky, A.T., Head, J.W., Schaber, G.G., \& Strom, R.G.  1997, in Venus II,
ed.  S.W.~Bougher, D.M.~Hunten, \& R.J.~Phillips, (Tucson:  University of Arizona Press),
1047

\medskip \noindent Bullock, M.A., Grinspoon, D.H., \& Head, J.W. 1993, GRL, 20, 2147

\medskip \noindent Buratti, B.J., Hicks, M.D., \& Newburn, R.L., Jr., 1999, Nature, 397,
219

\medskip \noindent Chapman, C.R., Merline, W.J., Bierhaus, B., Brooks, S., \& The Galileo Imaging
Team  1998, LPSC, XXIX, \#1927

\medskip \noindent Croft, S.K., Kargel, J.S., Kirk, R.L., Moore, J.M., Schenk, P.M., \&
Strom, R.G.,  1995, in Neptune and Triton, ed.  D.P.  Cruikshank (Tucson:  University of
Arizona Press), 879

\medskip \noindent Head, J.W., Crumpler, L.S., Aubele, J.E., Guest, J.E., \& Saunders,
R.S.  1992, in Venus II, ed.  S.W.~Bougher, D.M.~Hunten, \& R.J.~Phillips, (Tucson:
University of Arizona Press), 13153

\medskip \noindent Davis, D.R. and Farinella, P. 1997, it Icarus, 125, 50

\medskip \noindent Duncan, M.J., \& Levison, H.F.  1998, Science, 276, 1670

\medskip \noindent Farinella, P., Davis, D.R., \& Stern, S.A. 2000, in Protostars
and Planets IV, ed. V.~Mannings. (Tucson:  University of Arizona Press), in press

\medskip \noindent Goldreich, P., Murray, N., Longaretti, P.Y., \& Banfield, D.  1989,
Science, 245, 500

\medskip \noindent Holman, M.J., \& Wisdom, J.  1993, AJ, 105, 1987

\medskip \noindent Kirk, R.L., Soderblom, L.A., Brown, R.H., Kieffer, S.W., \& Kargel, J.S.
1995, in Neptune and Triton, ed.  D.P.~Cruikshank (Tucson:  University of Arizona Press),
849

\medskip \noindent Levison, H.F.  1997, in Completing the Inventory of the Solar System,
ed.  T.W.  Rettig \& J.M.  Hahn (San Francisco:  ASP), 173

\medskip \noindent Levison, H.F., \& Duncan, M.J.  1993, ApJ, 406, L35

\medskip \noindent --------.  1997, Icarus, 127, 13 (LD97)

\medskip \noindent Levison, H.F., Duncan, M.J., Zahnle, K., \& Dones, L.  1999, Icarus,
submitted

\medskip \noindent Lunine, J.I.  \& Nolan, M.  1992, Icarus, 100, 221 

\medskip \noindent McKinnon, W.B., \& Chapman, C.R.  1986, in Satellites, ed.
J.A.~Burns \& M.S.  Matthews (Tucson:  University of Arizona Press), 492

\medskip \noindent McKinnon, W.B.  1997, Nature, 390, 23

\medskip \noindent McKinnon, W.B., \& Leith, A.C.  1995, Icarus, 118, 392

\medskip \noindent McKinnon, W.B., \& Kirk, R.L.  1999, in Encyclopedia of the Solar
System, ed.  P.R.~Weissman, L.-A.~McFadden, \& T.V.~Johnson (San Diego:  Academic Press),
405

\medskip \noindent McKinnon, W.B., \& Schenk, P.M.  1995, GRL, 22, 1829

\medskip \noindent McKinnon, W.B., Lunine, J.I., \& Banfield, D.  1995, in Neptune and
Triton, ed.  D.P.  Cruikshank (Tucson:  University of Arizona Press), 807

\medskip \noindent Melosh, H.J.  1989, Impact Cratering:  A Geologic Process (New York:
Oxford University Press)

\medskip \noindent Neukum, G., Wagner, R., Wolf, U., Ivanov, B.A., Head, J.W., Pappalardo, R.T.,
Klemaszewski, J.E., Greeley, R., \& Belton, M.J.S.  1998, LPSC, XXIX, \#1742

\medskip \noindent Schenk, P.M.  1991, JGR, 96, 15635

\medskip \noindent --------.  1992, in Lunar and Planetary Science XXIII (Houston:  Lunar
Planet.  Inst.), 1215

\medskip \noindent Shoemaker, E.M.  1996, in Europa Ocean Conference Abstracts (San Juan
Capistrano Res.  Inst.), 65

\medskip \noindent Shoemaker, E.M., \& Wolfe, R.F.  1982, in Satellites of Jupiter, ed.
D.~Morrison (Tucson:  University of Arizona Press), 277

\medskip \noindent Smith, B.A., \& the Voyager Imaging Team 1989, Science, 246, 1422

\medskip \noindent Spencer, J.R., \& Schneider, N.M.  1996, ARE\&PS, 24, 125

\medskip \noindent Stern, S.A. 1995, AJ, 110, 856

\medskip \noindent Stern, S.A., \& McKinnon, W.B., 1999, LPSC, XXX, \#1766

\medskip \noindent Stevenson, D.J., \& Gahndi, A.S.  1990, in Lunar and Planetary Science XXI
(Houston:  Lunar Planet.  Inst.), 1202

\medskip \noindent Strom, R.G., Croft, S.K., \& Boyce, J.M.  1990, Science, 250, 437

\medskip \noindent Strobel, D.F., \& Summers, M.E.  1995, in Neptune and Triton, ed.  D.P.
Cruikshank (Tucson:  University of Arizona Press), 991

\medskip \noindent Weissman, P.R., \& Levison, H.F.  1997, in Pluto and Charon, ed.
S.A.~Stern \& D.J.~Tholen (Tucson:  University of Arizona Press), 559

\medskip \noindent Weissman, P.R., \& Stern, S.A.  1994, Icarus, 111, 378

\medskip \noindent Yelle, R.V., Lunine, J.I., Pollack, J.B., \& Brown 1995, in Neptune and
Triton, ed.  D.P.  Cruikshank (Tucson:  University of Arizona Press), 1107

\medskip \noindent Zahnle, K., Dones, L., \& Levison, H.F.  1998, Icarus, 136, 202

\vfill \eject

\centerline {FIGURE CAPTIONS}

\centerline { }

\noindent FIG.~1.---Crater diameter estimates for Triton from equations~(1) and (2), as a
function of both impactor diameter and velocity.  We take $g$=78 cm s$^{-2}$ for Triton.  We
take a minimum impactor velocity $v_{min}$=2.3 km s$^{-1}$, as set by the root-sum-square of
Triton's escape speed and the difference between Triton's orbital speed and the escape speed
from Triton's orbit.  We take a maximum impactor velocity $v_{max}$=11.6 km s$^{-1}$, as set
by the root-sum-square of Triton's escape speed and the sum of Triton's orbital speed and the
maximum impactor velocity at Triton's orbit (which is the root-sum-square of the escape speed
from Triton's orbit and, from LD97, the maximum encounter speed at Neptune's sphere of
influence).  Because Triton's surface is icy (e.g., Croft et al.~1995; McKinnon \& Kirk
1999), we assume values of $A$ and $\alpha$ appropriate for water ice, i.e., 0.20 and 0.65,
respectively (McKinnon \& Schenk 1995).  Model results were computed from Equations (1) and
(2) assuming a uniform distribution of impact velocities between $v_{min}$ and $v_{max}$.
The subtle upward curvature in the impactor size vs.~crater size is due to the diameter
correction for complex craters given in Equation (2b).  The four panels show various cases of
($\delta$,$\rho$) that bound the probable range of uncertainty with respect to Triton and
cometary impactors.  The two bold, horizontal lines represent the smallest crater size
counted and the largest crater seen on Triton, respectively (Strom, Croft, \& Boyce 1990).

\medskip \noindent FIG.~2.---Comparison of the differential size-frequency crater
distribution in Area 1 on Triton (solid black line) to model cases with varying impactor
size differential distribution power law slopes $b$ (green=--2.0, red=--2.5, and
blue=--3.0).  In all these model cases the minimum crater diameters shown are for $D$=2.8
km, which matches the smallest crater sizes counted in Area 1 (Strom, Croft, \& Boyce
1990).  To define absolute impact rates for this simulation, we normalized each model
case to the integral number of craters in the Area 1 dataset (181).  The four panels
display the same suite of four ($\delta$,$\rho$) cases as in Figure 1.

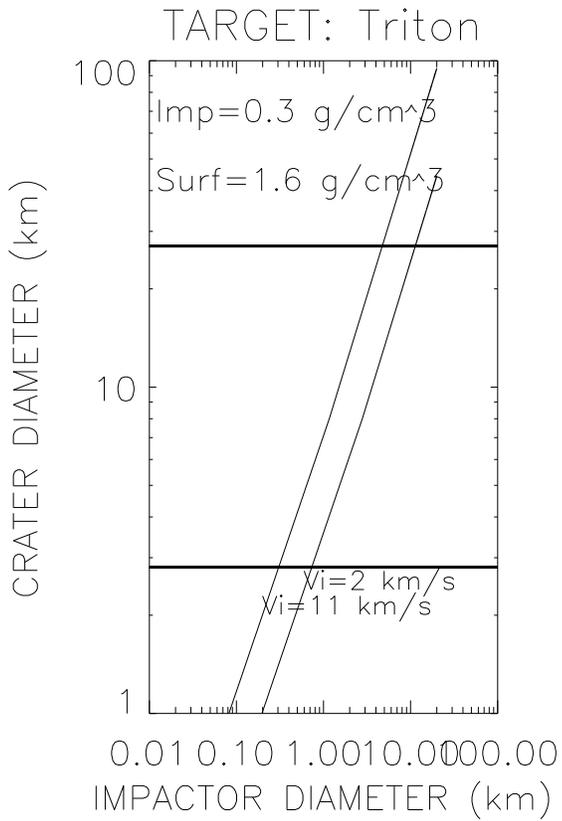
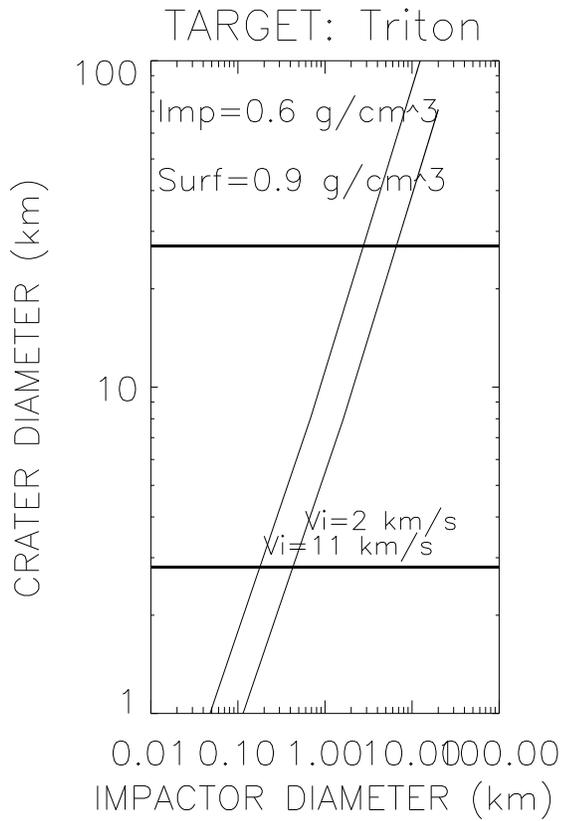
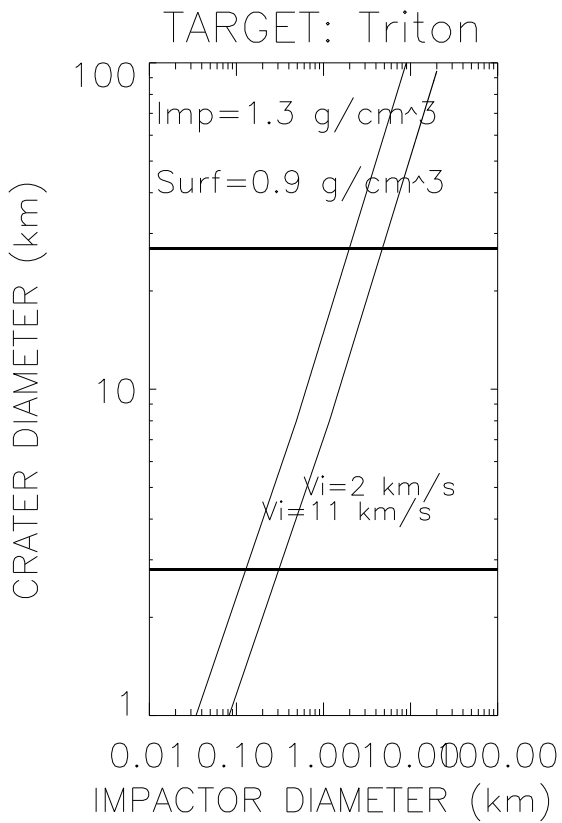
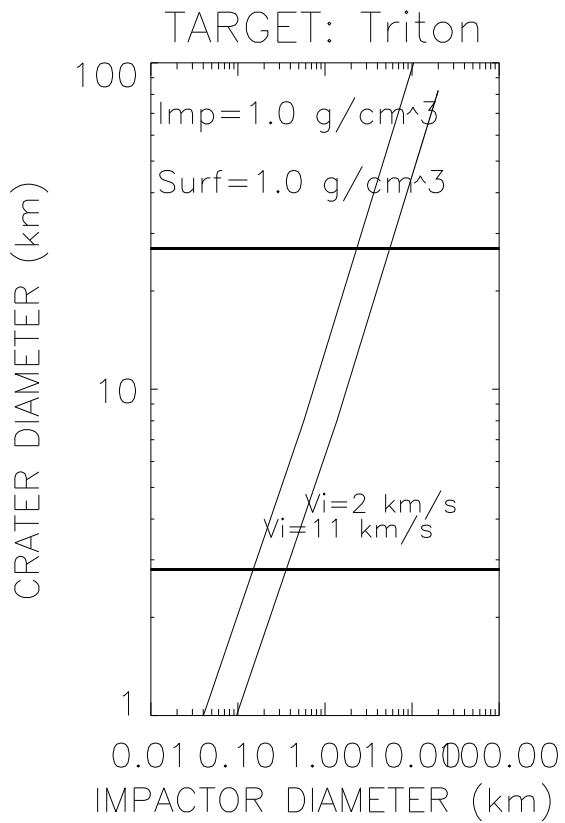
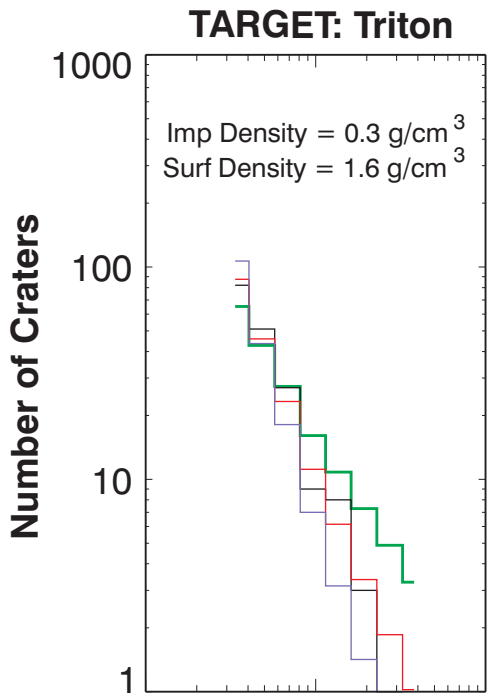
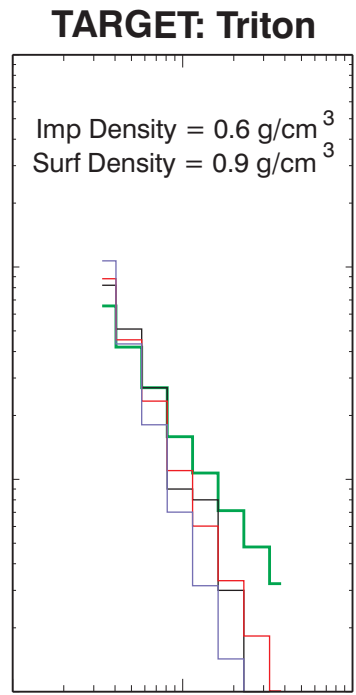
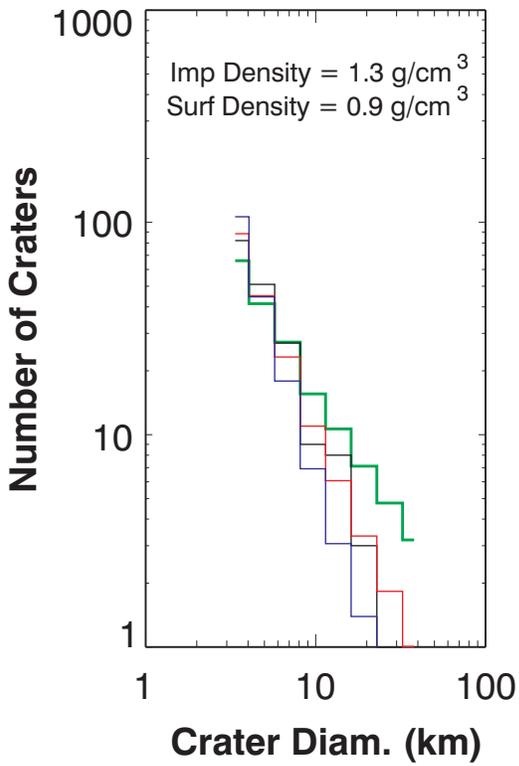
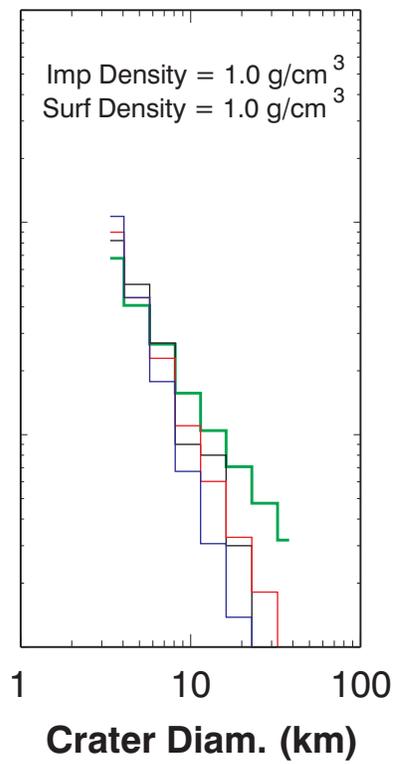
\bye